\documentclass[twocolumn]{aa}
  \usepackage{ifthen}
  \usepackage{amscd}
  \usepackage{amsmath}
  \usepackage{amssymb}
  \usepackage[T1]{fontenc}
  \usepackage[varg]{txfonts}
  \ifx\pdftexversion\undefined
    \usepackage[dvips]{graphicx}
  \else
    \usepackage[pdftex]{graphicx}
    \usepackage{epstopdf}
  \fi
  \usepackage{natbib}
  \usepackage{verbatim}
  \usepackage{array}

  \def\nar{New A Rev.}

\begin{document}

\title{Multi-spot ignition in type Ia supernova models}

\author{F. K. R{\"o}pke\inst{1}
        \and
        W. Hillebrandt\inst{1}
        \and
        J. C. Niemeyer\inst{2}
        \and
        S. E. Woosley\inst{3}}     
   \offprints{F. K. R{\"o}pke}

   \institute{Max-Planck-Institut f\"ur Astrophysik,
              Karl-Schwarzschild-Str. 1, D-85741 Garching, Germany\\
              \email{[fritz;wfh]@mpa-garching.mpg.de}
              \and
              Universit\"at W\"urzburg, Am Hubland, D-97074 W\"urzburg,
              Germany\\
              \email{niemeyer@astro.uni-wuerzburg.de}
              \and
              Department of Astronomy and Astrophysics, University of
              California, Santa Cruz, CA 95064, USA\\
              \email{woosley@ucolick.org}
             }

\abstract{
  We present a systematic survey of the capabilities of type Ia
  supernova explosion models starting from a number of flame seeds
  distributed around the center of the white dwarf star. To this end
  we greatly improved the resolution of the numerical simulations in
  the initial stages. This novel numerical approach facilitates a
  detailed study of multi-spot ignition scenarios with up to hundreds
  of ignition sparks. Two-dimensional simulations are shown to be
  inappropriate to study the effects of initial flame
  configurations. Based on a set of three-dimensional models, we
  conclude that multi-spot ignition
  scenarios may improve type Ia supernova models towards
  better agreement with observations. The achievable effect
  reaches a maximum at a limited number of flame ignition kernels as
  shown by the numerical models and corroborated by a simple
  dimensional analysis.
}

\maketitle


\section{Introduction}
\label{intro_sect}

Over the past years, a consensus has emerged about the general astrophysical
scenario of the majority of type Ia supernovae (SNe Ia).
These events are associated with thermonuclear explosions
of white dwarf (WD) stars close to the Chandrasekhar mass \citep[for a
recent review see][]{hillebrandt2000a}. A thermonuclear flame ignited
near the WD's center is believed to propagate outward
incinerating most parts of it.  The released nuclear energy
may suffice to explode the star as could be shown in several numerical
simulations \citep[e.g.][]{nomoto1984a, reinecke2002c, gamezo2003a,
  roepke2005b}. 

Yet little is known about the way the thermonuclear
flame ignites.
The evolution towards flame ignition is a complex physical process. As
a single WD is an inert object, dynamics must be introduced
into the progenitor system by assuming it to be a binary. The favored
scenario suggests a non-degenerate binary
companion from which the WD accretes matter. Due to this mass
accumulation it approaches the Chandrasekhar limit, steadily
increasing its central density so that eventually carbon burning
ignites. In the following several
hundreds of years a simmering phase of convective burning sets the
conditions under which finally a thermonuclear runaway occurs
leading to the the formation of a flame. 
Unfortunately, the flame ignition is difficult to address both
analytically and numerically. The few studies that tackled
the pre-ignition evolution proposed different scenarios for
flame formation.
While one study claims the flame ignition to take place in only
one single point near the center \citep{hoeflich2002a}, others favor
an ignition in multiple sparks distributed around the center or on
only one side of it, depending on the large-scale convective flow
pattern \citep{garcia1995a, woosley2004a, wunsch2004a,
  iapichino_phd}.

Not being well constrained by theory, shape and location of the
first flame(s) are usually
treated as free initial parameters in multi-dimensional simulations.
The hope was originally that those had only little impact
on the models supporting the observational finding of a remarkable
uniformity of SN Ia characteristics. However, multi-dimensional
simulations starting with different initial flame
configurations  disagreed with this conjecture.
The way of flame ignition turned out to be one of the influential
parameters of the models.
Different choices gave rise to controversial results. Single-spot
central ignitions leading to explosions of the WD were studied by
\citet{niemeyer1995b}, \citet{reinecke1999b}, \citet{hillebrandt2000b}, \citet{reinecke2002b},
\citet{reinecke2002c}, and
\citet{gamezo2003a}. In contrast, off-center
single-spot ignitions release insufficient energy to explode the
star \citep{niemeyer1996a, calder2004a}. Multi-spot
ignition scenarios were applied by 
\citet{niemeyer1996a}, \citet{reinecke2002d}, and \citet{roepke2005b}
and seem to have the potential to increase the explosion strength. 
Apparently, a slight misalignment of the initial flame with the WD's
center leads to a drastic change in the outcome of the simulations
when starting with a single perfect sphere \citep{calder2004a}. More
structured multi-spot flame configurations are less sensitive to it
\citep{roepke2005b}. But even setting aside the complications arising
from asymmetries in off-center ignitions, the results of the models
have been shown to depend on the number of flame seeds
\citep{reinecke2002d, travaglio2004a}.
The objective of the present study is to explore this effect in
detail in a systematic approach.

The question of what can be expected from multi-spot ignition
scenarios with an increasing number and different distributions of
initial flames was addressed only recently by \citet{garcia2005a}. We
report on a similar study, which is, however,
based on a completely different approach to modeling thermonuclear
supernovae. While \citet{garcia2005a} apply a Lagrangian description
of the hydrodynamics based on the smooth particle hydrodynamics (SPH)
technique, we utilize an Eulerian grid-based finite volume method. The
key distinction is that
our approach facilitates a self-consistent description
of turbulent thermonuclear flame propagation, as will be
discussed in Sect.~\ref{nummeth_sect}. In contrast, the SPH model
has to rely on a 
parameterization of the effective turbulent flame propagation
velocity, since it cannot provide a valid description of turbulence
effects.
The predictive power of the SPH approach is thus limited. Nevertheless, the
large-scale structures observed by \citet{garcia2005a}
appear to be similar to what we find
in our simulations, as can be seen from a comparison with a
full-star model presented by \citet{roepke2005b}. This is not too
surprising since both are driven by buoyancy and should be equally
well reproduced in both approaches. 

As will be
discussed below, the simulation parameters and the details of the
results of the present study differ significantly from those of
\cite{garcia2005a}. Moreover, the focuses of the two surveys are
distinct. While \cite{garcia2005a} presented the first attempt to modeling
SNe Ia in three dimensions (albeit in a parametrized way) from flame
formation on and to assess
the resulting explosions, we take a more pragmatic point of view. Our
objective is to answer the question
whether multi-spot ignition scenarios are in principle capable of
curing some of the shortcomings of current deflagration SN Ia models.
We therefore apply a number of different initial flame configurations
without attempting to model their pre-ignition evolution. The
question of how realistic these are is set aside in the present study.

Although being within the range of observational expectations, the
explosion energies and the masses of burning products in all
multi-dimensional deflagration models presented so far seem to be on
the weak side and multi-spot ignition scenarios may offer a way to
improve the results.  
Besides such global quantities, we are especially
interested in the distribution of the species in velocity space. 
Previous work showed that poorly resolved models 
are inconsistent with observations in predicting 
low-velocity oxygen and carbon lines in
late time spectra \citep{kozma2005a}. This is due to downdrafts of
unburnt material in between burning buoyancy-driven bubbles of
ashes. These downdrafts transport significant amounts of carbon and
oxygen towards the center of the WD. It may well be, however, that
the downdrafts carry sparks seeding additional burning, which are not
resolved in current simulations.
Moreover, the simulation from
which the synthetic spectrum of \citet{kozma2005a} was derived started
out with a very artificial initial central flame
configuration. The model was calculated on a uniform
computational grid of  $[256]^3$ cells co-expanding with the WD. This
led to an only marginally converged result and the explosion energy as
well as the production of iron group and intermediate mass elements
was rather low. Obviously, some of the disagreements with the
observations are caused by the simplicity of the model -- in particular
the initial flame shape -- and one may wonder whether the
problems are mitigated with more structured flame configurations such
as arising from multi-spot ignition.

An estimate of the capabilities of the multi-spot ignition scenario is
given in Sect.~\ref{multi_sect}.
In Sect.~\ref{nummeth_sect} we will briefly outline the
techniques underlying our numerical simulations and discuss a new
implementation that enables us to study multi-spot ignition scenarios
in detail. This implementation is compared with previous simulations
in Sect.~\ref{test_sect}. 
Results from two-dimensional models will
be discussed in Sect.~\ref{twod_sect}. Although it will be shown that
these are inappropriate to study multi-spot ignitions, they point to
important effects. A systematic
survey based on three-dimensional simulations is presented in
Sect.~\ref{threed_sect}. In Sect.~\ref{concl_sect} we draw conclusions
for modeling thermonuclear supernova explosions.

\section{Multi-spot ignition: estimating the gains}
\label{multi_sect}

\citet{woosley2004a} and \citet{wunsch2004a} conclude from analytical
models of the ignition process that a multi-spot ignition is possible
(but not guaranteed). Timescale arguments show that a number of hot
spots is in principle capable of evolving towards a thermonuclear
runaway. The first flame will ignite at a radius $\sim$$150 \,
\mathrm{km}$ off-center of the WD and more can follow. Yet their
number and spatial distribution cannot be conclusively constrained by
the analytical models.

Thus, it is justified to regard these as free parameters in
simulations of the explosion process. In this spirit, we address
the question of the number of ignition spots in the present
study. Concerning the spatial distribution we simplify matters by
assuming a spherically symmetric probability density of ignitions to
occur around the WD's center. This ignores possible large-scale
anisotropies of the ignition process due to low-order modes in the
convective flow pattern prior to ignition \citep{woosley2004a}. The
impact of such effects is subject to a forthcoming publication.

Given the ambivalence of the ignition conditions, it is a legitimate
question to ask which multi-spot configuration would lead to an
optimal fuel consumption producing the most vigorous explosion and
potentially burning most of the fuel near the center. A
dimensional analysis such as presented in the following can shed some
light on this issue. Sects.~\ref{twod_sect} and \ref{threed_sect} will
tackle the question from the side of numerical simulations.

One has to note that there are two major effects on the outcome
of multi-spot ignitited supernova simulations. If a single flame seed is
separated from the bulk of ignition sparks towards larger radii, it
will experience a larger gravitational acceleration. The burning front
resulting from this spark will evolve faster than the other ignition
points and dominate the flame evolution. This effect is determined by the
nonlinear evolution of the burning fronts and hard to predict
analytically. It will thus be discussed on the basis of numerical
simulations in Sects.~\ref{twod_sect} and \ref{threed_sect}. Here, we
only note that the distribution of ignition points should not be too
sparse in order to avoid such effects. 

On the other hand, accomodating
too many ignition kernels in a given volume will have the effect
that the flames merge shortly after ignition due to
self-propagation. In this case, the
simulation will look similar to those ignited centrally in a single
connected shape. The main
advantage of multi-spot ignition scenarios, i.e.\ a large flame
surface, is lost. Two effects counteract the surface distruction by
merging of the fronts. Firstly, the overall expansion of the star due
to the nuclear energy release leads to an increasing separation of
burning bubbles and, secondly, buoyancy-induced flotation rises them
in radial direction further increasing their separation.

In a simplified picture,
two igniting bubbles of radius $r_\mathrm{b}$ at a separation $l$ will
lead to a maximum flame surface if their growth due to flame
propagation is exactly compensated by these two effects.
Assuming a homologous expansion of the star, the temporal change in
distance of the bubbles, $\dot{l}_\mathrm{ex}$ induced by this
effect is given by  
\begin{equation}
\frac{l - 2r_\mathrm{b}}{\dot{l}_\mathrm{ex}} = \tau_\mathrm{dyn},
\end{equation}
where $\tau_\mathrm{dyn} \sim (G \rho)^{-1/2} = 0.07\, \mathrm{s}$ is
the dynamical timescale assuming a density of $3 \times 10^9 \,
\mathrm{g} \, \mathrm{cm}^{-3}$ ($G$ denotes Newton's constant).

Due to buoyancy, two neighboring bubbles enclosing an angle $\phi$ with
the star's center will rise radially, increasing their separation by
\begin{equation}
\dot{l}_\mathrm{buoy} = 2 \dot{r} \sin \frac{\phi}{2}.
\end{equation}
For the angle $\phi$ we assume
\begin{equation}
\sin \frac{\phi}{2} = \frac{l - 2r_\mathrm{b}}{2 \langle r \rangle},
\end{equation}
with an average initial distance $\langle r \rangle$ of the bubbles
from the WD's center. Thus,
\begin{equation}
\dot{l}_\mathrm{buoy} = \dot{r} \frac{l- 2r_\mathrm{b}}{\langle r
  \rangle}
\end{equation}
The asymptotic radial rise velocity of a bubble is (up to a factor of order
unity) given by
\begin{equation}
\label{vrt}
\dot{r} = \sqrt{\mathrm{At}\, g \, r_\mathrm{b}}
\end{equation}
\citep{davies1950a}, where $\mathrm{At} = (\rho_\mathrm{u} - \rho_\mathrm{b}) /
(\rho_\mathrm{u} + \rho_\mathrm{b}) $ denotes the Atwood number across
the flame front with the density $\rho_\mathrm{u}$ of unburnt material
outside and the density $\rho_\mathrm{b}$ of the burnt material inside
the bubble. In its functional form, Eq.~(\ref{vrt}) follows the
  expression obtained from balancing the buoyancy and the drag forces
  for a spherical bubble. Typical values are $g \sim 10^9 \,
\mathrm{cm} \, \mathrm{s}^{-2}$ and  about 0.07 for the Atwood number at a
fuel density of $\rho_\mathrm{b} \sim 3
\times 10^9 \, \mathrm{g} \, \mathrm{cm}^{-3}$\citep{timmes1992a}.

The distance decrement by growth of the bubbles due to burning is
given by the propagation velocity of the flame front,
\begin{equation}
\dot{l}_\mathrm{burn} = 2 v_\mathrm{burn},
\end{equation}
with $v_\mathrm{burn} \sim 10^7 \, \mathrm{cm}\,\mathrm{s}^{-1}$
shortly after ignition.
Equating the sum of $\dot{l}_\mathrm{ex}$ and $\dot{l}_\mathrm{buoy}$
with $\dot{l}_\mathrm{burn}$ yields for
the optimal initial bubble separation $l_\mathrm{opt}$
\begin{equation}
l_\mathrm{opt} - 2r_\mathrm{b} = \frac{2 v_\mathrm{burn} \tau_\mathrm{dyn}
\langle r \rangle}{\langle r \rangle  + \tau_\mathrm{dyn}
\sqrt{\mathrm{At}\, g \, r_\mathrm{b}}}.
\end{equation}
The average density of bubbles $n$ in a sphere of radius $R$ is given by
\begin{equation}
n = \frac{3 N}{4 \pi R^3},
\end{equation}
with $N$ denoting the number of bubbles. In the optimal case, this
should be equal to the inverse volume of a sphere with radius
$0.5 l_\mathrm{opt} + r_\mathrm{b}$, i.e.\ the number of bubbles in
this case is given by
\begin{equation}
\label{n_opt_eq}
N_\mathrm{opt} = \left\{ \frac{R \,\left(\langle r \rangle +
    \tau_\mathrm{dyn} \sqrt{\mathrm{At}\, g \, r_\mathrm{b}}\right)}
{\tau_\mathrm{dyn} v_\mathrm{burn} \langle r \rangle + 2 r_\mathrm{b}
  \langle r \rangle + 2 \tau_\mathrm{dyn} r_\mathrm{b}
  \sqrt{\mathrm{At}\, g \, r_\mathrm{b}}
}
\right\}^3.
\end{equation}
Thus, assuming a central density of the WD of $3 \times 10^9
\,\mathrm{g}\,\mathrm{cm}^{-3}$, an ignition radius of 100 to 150$\,
\mathrm{km}$ \citep{woosley2004a}, and setting the average initial
distance from the 
center $\langle r \rangle$ to one half of this radius, the optimal number
of ignition sparks with a
radius of $r_\mathrm{b} = 5.5 \, \mathrm{km}$ (being the smallest
resolvable bubble structure in the numerical models discussed below)
should be somewhere between 23 and 76 per octant.
This, of course, is
no more than a crude estimate. Due to burning the bubbles will increase
in size and this in turn affects the energy release and the
contribution from buoyancy effects.
Moreover, the establishing flows
and the interaction of the flame with turbulence have been
ignored. In
a realistic numerical simulation, the flame seed configuration is much
more complex. In most setups (see Sects.~\ref{twod_sect} and
\ref{threed_sect}) we will apply a Gaussian distribution in radius.
Since the outer bubbles are more separated in this case, the optimal
number of bubbles is expected to be larger than the simple estimate.

\begin{figure}[t]
\centerline{
\includegraphics[width = \linewidth]
  {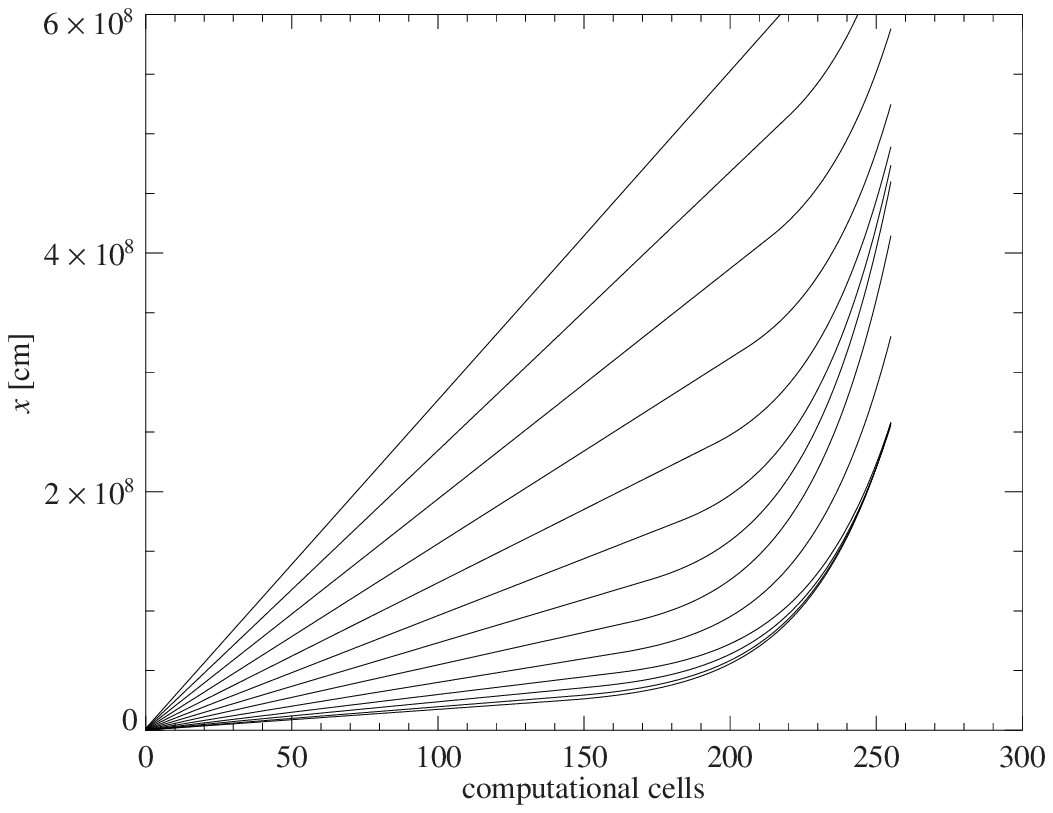}}
\caption{Evolution of the hybrid computational grid in a
  three-dimensional simulation. Each curve corresponds to a time
  progression of $\Delta t = 0.1 \, \mathrm{s}$, starting at the
  lowest curve with $t = 0 \, \mathrm{s}$.
  \label{gridevo_fig}}
\end{figure}

\begin{figure*}[t]
\centerline{
\includegraphics[width = \linewidth]
  {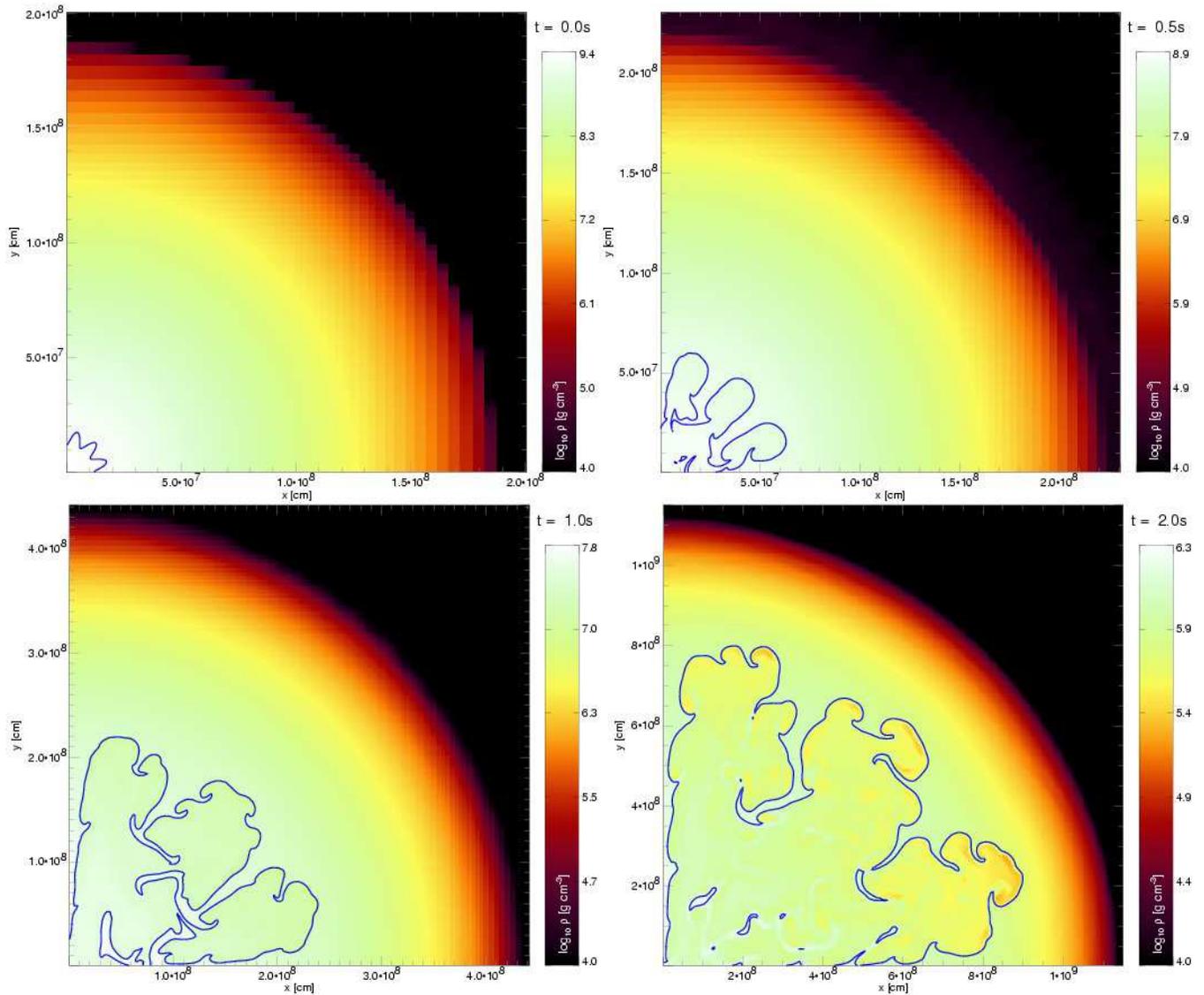}}
\caption{Snapshots of the evolution of a \emph{c3}-model on a $256^2$
  cells hybrid grid. The logarithm of the density is color-coded and the
  contour marks the location of the flame front.
  \label{2d_c3_evo}}
\end{figure*}

With Eq.~(\ref{n_opt_eq}) the question can be answered of
whether it is possible to tune models to give more vigorous explosions
by increasing the
resolution and accommodating an ever larger number of ignition
spots. Even in the limit of $r_\mathrm{b} \rightarrow 0$ the optimal
number $N_\mathrm{opt}$ of flame seeds increases to no more than 354
to 1195 per octant. Thus the effect of the number of initial flame
sparks on the explosion strength is limited.

\section{Explosion: astrophysical scenario and numerical methods}
\label{nummeth_sect}

The explosion model describes the propagation of
the flame after ignition near the center of the WD outwards. Our
implementation is based on the deflagration scenario in which the
flame velocity is 
sub-sonic and accelerated by the interaction with turbulence generated
by generic instabilities \citep[see e.g.][]{hillebrandt2000a}. It
follows the numerical methods developed by \citet{reinecke1999a,
  reinecke2002b} and \citet{roepke2005c}. For the details and
extensive tests of the
implementation we refer to these works. The fundamental concept
behind the approach is that of large eddy simulations (LES). Only the
largest scales of the problem are directly resolved. The
interaction of the flame with turbulent eddies on unresolved scales of
the establishing turbulent energy cascade is described by a sub-grid
scale turbulence model \citep{niemeyer1995b}. The large separation of
the flame width (of the order of a millimeter) from the resolved scales
justifies the flame modeling as a sharp discontinuity separating the
fuel from the ashes. To this end, the level-set technique is applied
as described in detail by \citet{reinecke1999a}.

We apply the same WD equation of state as used by
\cite{reinecke2002b}. In all simulations presented in the present
paper we fix the initial composition to equal parts of carbon and
oxygen throughout the star and construct the initial WD assuming cold
isothermal ($T = 5 \times 10^5 \, \mathrm{K}$) conditions and a central
density of $2.9 \times 10^9 \, \mathrm{g} \, \mathrm{cm}^{-3}$. The
nuclear reactions are implemented in the simplified approach suggested
by \citet{reinecke2002b}. Five species ($^{12}$C, $^{16}$O, $^{24}$Mg
representing the intermediate mass elements,
$^{56}$Ni as a representative of the iron group, and
$\alpha$-particles) are followed\footnote{We will set ``Ni'' and
  ``Mg'' in quotes henceforth to avoid confusion with the actual
  isotopes.}.
At high fuel densities the material
crossed by the
flame is converted to nuclear statistical equilibrium modeled as a
mixture of ``Ni'' and $\alpha$-particles. Below fuel densities of
$5.25 \times 10^7 \, \mathrm{g} \, \mathrm{cm}^{-3}$ intermediate mass
elements are produced and once the fuel density drops below $10^7 \,
\mathrm{g} \, \mathrm{cm}^{-3}$ we stop burning. In the current
implementation electron captures are neglected. These could decrease the
density in the ashes and therefore increase the density contrast over
the flame. In the setups chosen for our simulations we expect the
effect on the dynamics of the explosion to be small,
but significant changes may result at still higher central densities
than those applied here.

Our novel approach that enables us to study the effects of initial
flame configurations in detail is a
computational grid different from previous implementations. While the
first three-dimensional simulations by \citet{hillebrandt2000b} and
\citet{reinecke2002b, reinecke2002d,reinecke2002c} were carried
out on static Cartesian grid geometries with a fine-resolved uniform
inner part and an exponentially growing grid spacing further out to
capture parts of the expansion in the explosion process,
\citet{roepke2005c} and \citet{roepke2005b} applied a moving uniform
grid that tracked the expansion of the WD star.
Here, we combine both approaches and use a moving grid that is
composed of two nested sub-grids. The inner part (\emph{grid 1}) again
features a uniform fine
resolution and the outer grid cells (\emph{grid 2}) grow
exponentially. Initially, both
grids are moved individually. While \emph{grid 1} contains the flame and
tracks its propagation, \emph{grid 2} follows the WD expansion.
Since in the course of the explosion the flame spreads out over the WD
star, \emph{grid 1} must expand faster than \emph{grid 2}. It is
therefore possible to subsequently gather cells of \emph{grid 2} into
\emph{grid 1} as soon as the grid spacings match. Eventually,
\emph{grid 1} incorporates \emph{grid 2} and the full WD is covered by
a single uniform grid tracking its expansion. We will call this
approach \emph{hybrid grid} in the following. An example for the grid
evolution in one of our simulations is given in
Fig.~\ref{gridevo_fig}.

Although conceptually simple, the hybrid grid has proven to be extremely
useful and is much easier to handle than adaptive mesh
refinement strategies. Since
the expansion and flame spread in the explosion is on average
spherical for most ignitions scenarios, it offers the
possibility to maximally resolve the flame region with a given fixed
number of computational cells. This leads to an improved modeling
of the flame propagation and, in particular, provides the possibility to
drastically improve the resolution in the central parts at the onset
of the explosion. In this way it becomes feasible to resolve highly
structured initial flame configurations. For studying multi-spot
ignition scenarios, the number of initial flame kernels could be
substantially increased. Whereas the previous static grid setups of one octant of
the WD star could resolve up
to $\sim$30 bubbles in a simulation with $768^3$ grid cells
\citep[cf.][]{travaglio2004a}, a 
hybrid-grid setup with $256^3$ cells can accommodate several hundreds of
ignition spots.

All simulations performed in this study are carried out on
only one spatial octant of the WD assuming mirror symmetry to the
other octants. As demonstrated by \citet{roepke2005b}, this symmetry
constraint does not suppress a development of potential low-wavenumber modes
in the nonlinear flow pattern. Although these may in principle occur in
convective phenomena, the short explosion time-scale in SNe Ia
prevents them. Therefore restricting the simulations to only one
octant does not miss physical effects and provides a valid approach
to save computer time in our study.

\section{Testing the implementation}
\label{test_sect}

\begin{figure}[t]
\centerline{
\includegraphics[width = \linewidth]
  {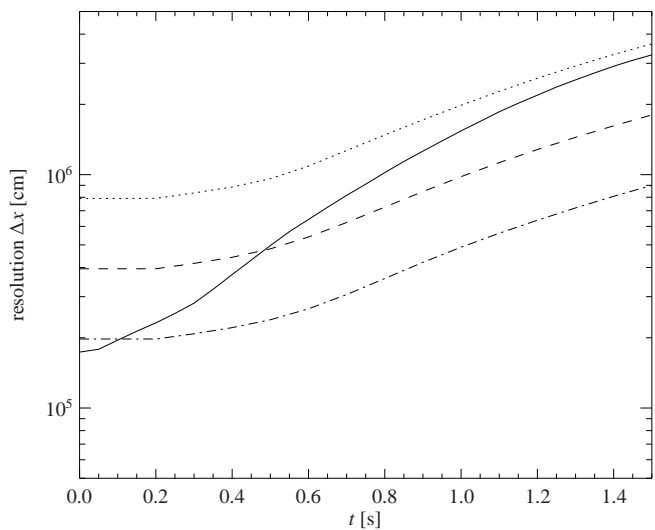}}
\caption{Flame resolution in different simulations:
$256^2$cells hybrid grid (solid), $256^2$cells uniform grid (dotted),
$512^2$cells uniform grid (dashed), and $1024^2$cells uniform grid
(dashed-dotted).
  \label{fl_res_fig}}
\end{figure}

Apart from a better resolution of the ignition, our
approach concentrates resolution in the flame region also in the
subsequent stages of the explosion. It is
therefore worthwhile to compare the results with previous
simulations. In particular, we will point out some changes with respect
to the expanding uniform grid simulations of \citet{roepke2005c}. For
this comparison, we apply the standard test case of a single-octant
simulation in which the flame is centrally ignited with a sinusoidal
(2d) or toroidal (3d) perturbation imposed on it. According to the
notation of \citet{reinecke2002b}, we will refer to this flame setup as
\emph{c3}.

Figure~\ref{2d_c3_evo} shows snapshots from the evolution of a
\emph{c3}-model in two dimensions imposing cylindrical
symmetry. Here, a $256^2$ cells hybrid grid
was applied. The initial \emph{c3} flame is depicted in the upper left
panel of
Fig.~\ref{2d_c3_evo}. Comparing these snapshots to those of the
previous uniform-grid implementations \citep[cf.\ Fig.~4
of][]{roepke2005c}, we note that the global structure resembles the
simulations  there
carried out on a much larger number of computational cells. These,
however, possess finer sub-structures. The reason is clear from a
comparison of the resolutions of the flame fronts in the different
implementations provided by Fig.~\ref{fl_res_fig}. Due to the moving
grids the resolution changes with time in all cases. As expected, the
concentration of computational cells in the flame part leads to a much
better resolution of the flame in the first stages for the hybrid
implementation on $256^2$ cells than for the uniformly expanding
$256^2$ cells grid. It is similar to the uniformly expanding $1024^2$
cells grid implementation. Therefore, the large-scale structures,
seeded by the initial \emph{c3} flame perturbations evolve in a
balanced way, whereas in the older implementation with $256^2$ grid
cells the inner Rayleigh-Taylor finger was suppressed \citep[cf.\ Fig.~4
of][]{roepke2005c}. Since the
flame propagates faster than the WD expands, the resolution in later
stages -- when the two nested grids have evolved into a single uniform
grid -- converges towards that of the previous
implementation. Naturally, this prevents the flame from developing as
detailed fine structures as the uniform grid $1024^2$
implementation. This effect is, however, compensated with regard to the
global quantities by the sub-grid scale turbulence model. We conclude
that our novel implementation provides a reasonable compromise between
computational expenses from larger grids and resolution of the flame.

\begin{figure}[t]
\centerline{
\includegraphics[width = 0.88 \linewidth]
  {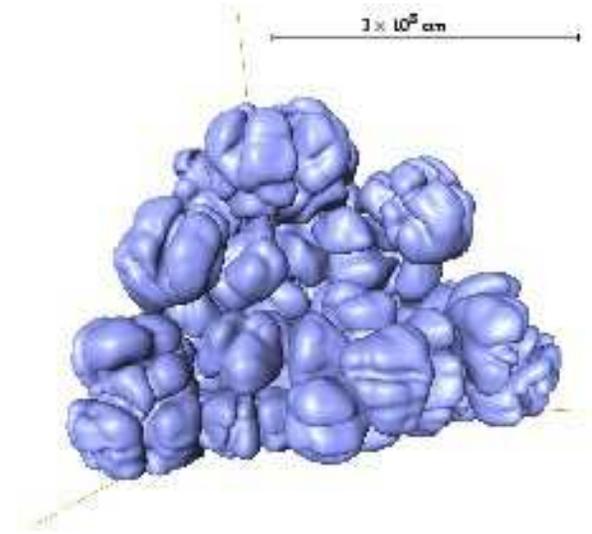}}
\caption{Flame front in model \emph{c3\_3d} $1.0 \, \mathrm{s}$ after ignition.
  \label{inimod_c3_fig}}
\end{figure}

The situation is very similar in three-dimensional simulations. A
snapshot from the standard \emph{c3\_3d} model $1.0 \, \mathrm{s}$
after ignition is shown in Fig.~\ref{inimod_c3_fig}. As in the
two-dimensional case, the three-dimensional implementation of the
hybrid grid leads to a more balanced evolution of the flame morphology
imprinted on the initial flame seed. The \emph{c3\_3d}
model in previous uniform grid implementations \citep[see Fig.~13 of
][]{roepke2005c} showed three strong flame features evolving along the
axes due to a suppression of initial perturbations. This
grid-imprinted symmetry is relaxed
in the novel implementation in which the flame is structured into more
and smaller features (see Fig.~\ref{inimod_c3_fig}). Interestingly,
the energy production is not increased by these effects in the
three-dimensional simulations \citep[cf. the energy release of the
\emph{c3\_3d} model on a uniform expanding grid presented by][]{roepke2005c}, corroborating
the conclusion that the description of the burning (at least in the
stages where it proceeds to iron group elements) is numerically
converged in our models. Although increased resolutions give rise to a modified
structure of the flame and a redistribution of the ashes, the
global quantities are largely unaffected.

\begin{figure*}[p]
\centerline{
\includegraphics[width = 0.95 \linewidth]
  {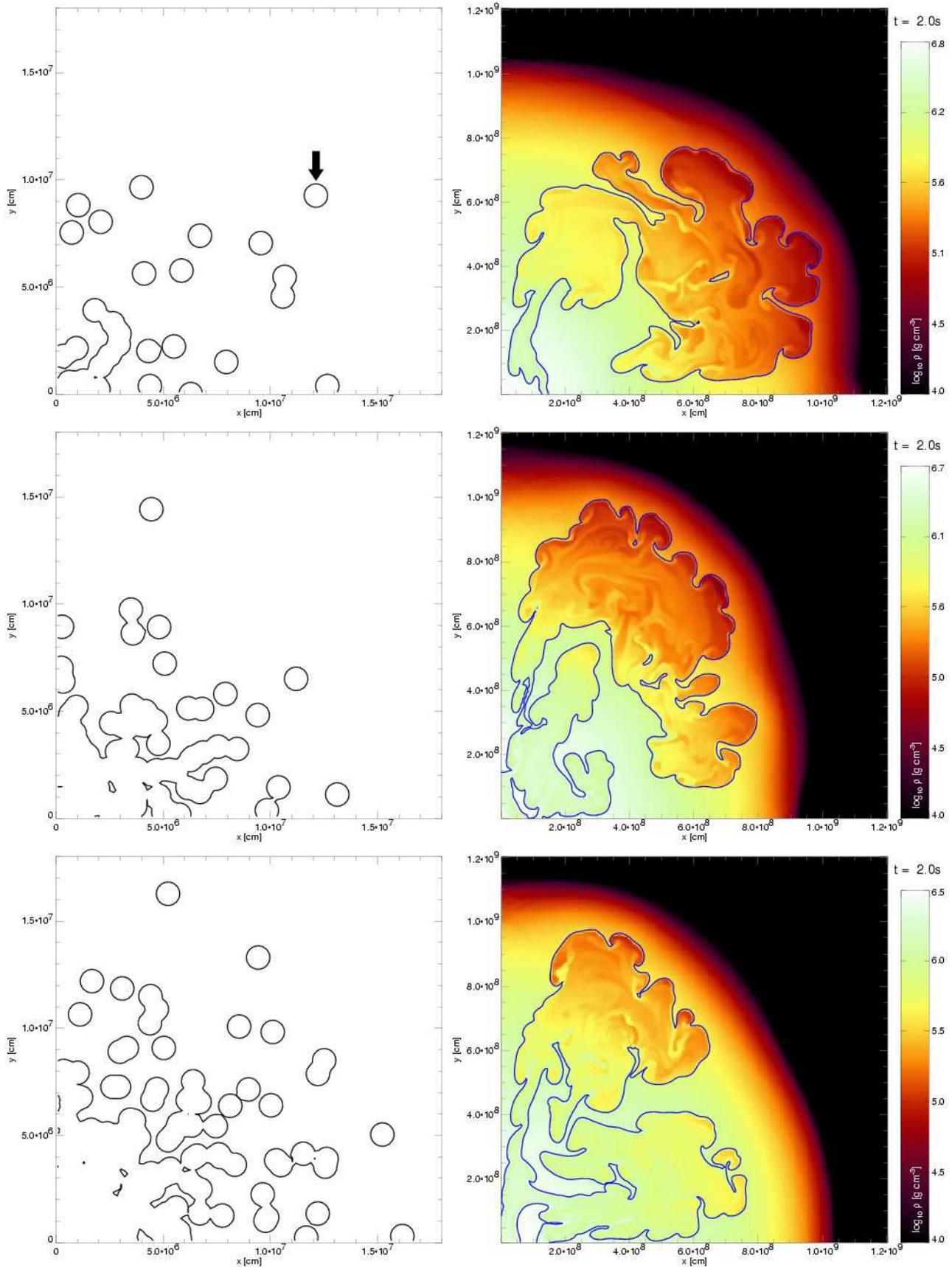}}
\caption{Two-dimensional multi-spot simulations: model \emph{b30\_2d}
  (top), model \emph{b100\_2d} (middle), and model \emph{b200\_2d}
  (bottom). The left column shows the initial flame configuration and
  the right column provides snapshots of the evolution after $2 \,
  \mathrm{s}$ with the the logarithm of the density color-coded and the flame
  front represented by the contour. The arrow in the top left panel marks the bubble that
  later on dominates the flame evolution.
  \label{2d_ms_fig}}
\end{figure*}

\section{Two-dimensional simulations}
\label{twod_sect}

With the hybrid grid implementation, four different flame
configurations were tested in two-dimensional simulations again
assuming a cylindrical symmetry. All were
carried out on $256^3$ grid cells with an initial inner grid spacing
of $1.74 \, \mathrm{km}$. Apart from
the \emph{c3\_2d}-model, different numbers of initial bubbles were chosen
in the following procedure. The bubbles with a radius of $5.5 \,
\mathrm{km}$ were distributed 
randomly in angle and
according to a Gaussian probability
distribution with a dispersion of $\sigma = 75 \, \mathrm{km}$
in radius. This Gaussian radial distribution,
however, was distorted by the constraint that no bubble was allowed to
ignite at a radius larger than $2.5 \sigma$ and by imposing a minimum
distance of the bubble centers $d_\mathrm{min}$. The models
\emph{b30\_2d}, \emph{b100\_2d}, and \emph{b200\_2d}
contained 30 bubbles with $d_\mathrm{min} = 0.8$, 100 bubbles
with $d_\mathrm{min} = 0.2$, and 200 bubbles with $d_\mathrm{min} =
0.1$, respectively. 

\begin{figure}[t]
\centerline{
\includegraphics[width = \linewidth]
  {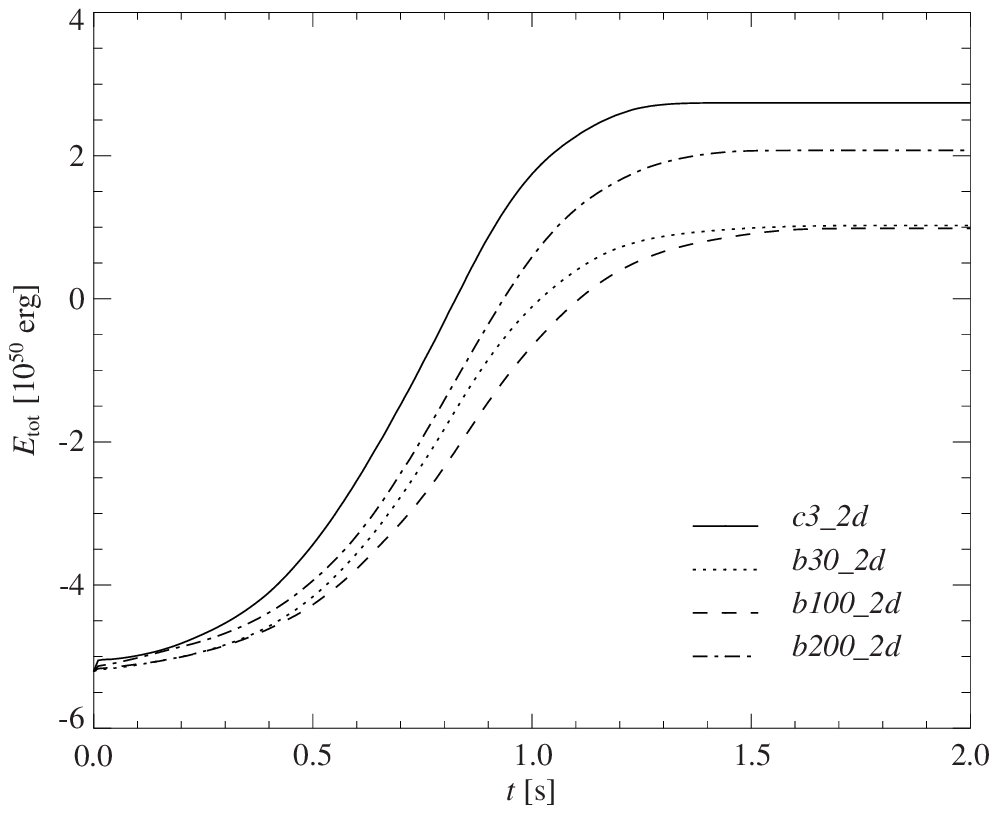}}
\caption{Total energies in the two-dimensional simulations.
  \label{2d_etot_fig}}
\end{figure}

The initial flame configurations are shown in
Fig.~\ref{2d_ms_fig}. This figure also contains snapshots at $t = 2 \,
\mathrm{s}$. From a comparison with the \emph{c3\_2d}-model in
Fig.~\ref{2d_c3_evo} it is clear that in the two-dimensional case
multi-spot ignition scenarios do not give rise to an increased overall fuel
consumption. This is confirmed by the evolution of the total
energies in the different models plotted in Fig.~\ref{2d_etot_fig}.
The models \emph{b30\_2d} and \emph{b100\_2d}
produce only about 40\% of the asymptotic kinetic energy of model
\emph{c3\_3d} and the energy
production of model \emph{b200\_2d} falls in between.

The reason for this finding is a peculiarity of two-dimensional
models. Apart from the fact that turbulence in the
two-dimensional case follows a different scaling than
three-dimensional turbulence, two-dimensional simulations tend to
amplify large structures. This may be attributed to the missing degree
of freedom in the third spatial direction where axial symmetry is
imposed. Thus the initial flames consist of tori rather than
bubbles. The large-scale flows around these structures are expected to
significantly differ from the flow patterns in three-dimensional
simulations. 

\begin{figure*}[t]
\centerline{
\includegraphics[width = \linewidth]
  {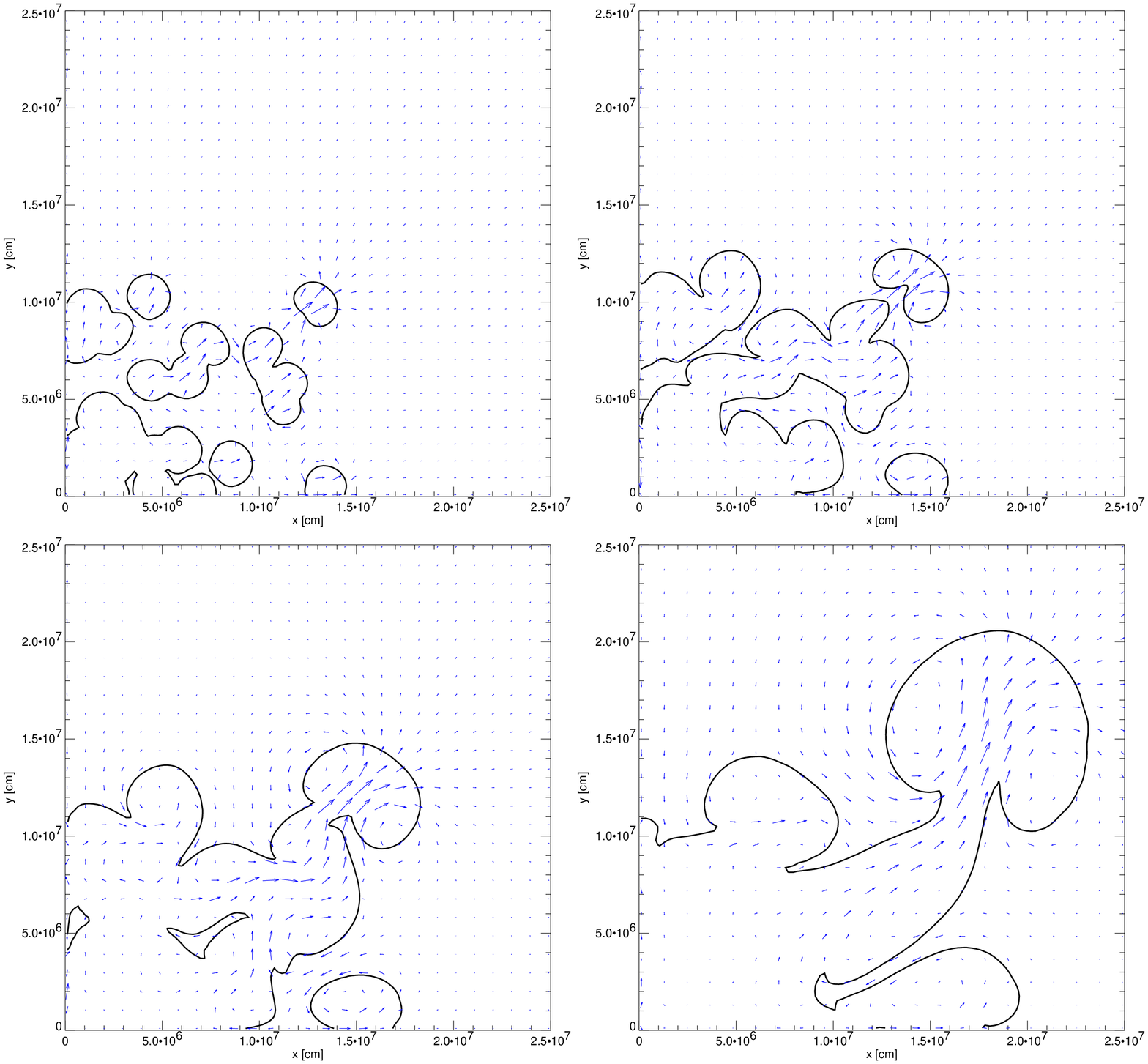}}
\caption{Flame evolution and velocity fields in model \emph{b30\_2d}
  at $t = 0.05 \, \mathrm{s}$, $t = 0.10 \, \mathrm{s}$,$t = 0.15 \,
  \mathrm{s}$, and $t = 0.25 \, \mathrm{s}$ (top left to bottom right).
  \label{b30_2d_evo_fig}}
\end{figure*}

As a result, in cases with few ignition spots, the outermost
flame seed dominates the evolution. This is illustrated by snapshots of the
early flame evolution in model \emph{b30\_2d} in
Fig.~\ref{b30_2d_evo_fig}. Starting from the initial configuration
shown in Fig.~\ref{2d_ms_fig}, the inner flames merge quickly, but the
outermost ignition spot (marked with the arrow in
Fig.~\ref{2d_ms_fig}) evolves into a structure that dominates the
flame evolution further on (cf.\ the snapshots in
Fig.~\ref{b30_2d_evo_fig} and the snapshot at $t = 2 \, \mathrm{s}$ in
Fig.~\ref{2d_ms_fig}). A similar effect governs the flame evolution in
model \emph{b100\_2d}. Due to the larger number of ignition spots in
model \emph{b200\_2d} the flame propagation is more balanced here and
finally becomes dominated by two large features (see Fig.~\ref{2d_ms_fig}).
The reason why the \emph{c3\_2d}-model appears so well-behaved is the
symmetric and balanced perturbation we impose on the initial flame. It
imprints three features which evolve equally.

Our results are in reasonable agreement with the findings of
\citet{niemeyer1996a} and the recent study by \citet{livne2005a}, both
reporting on dominating large-scale flame features. 

Thus our findings point to the effect mentioned in
Sect.~\ref{multi_sect}. Single spots may lead to
large-scale features which have an unfavorable effect on the
burning. Since flow patterns in two-dimensional simulations differ
significantly from those in three-dimensional cases, the question
arises of how pronounced this effect is there.

\section{Three-dimensional simulations}
\label{threed_sect}

\subsection{Ignition setups}

\begin{figure}[t]
\centerline{
\includegraphics[width = \linewidth]
  {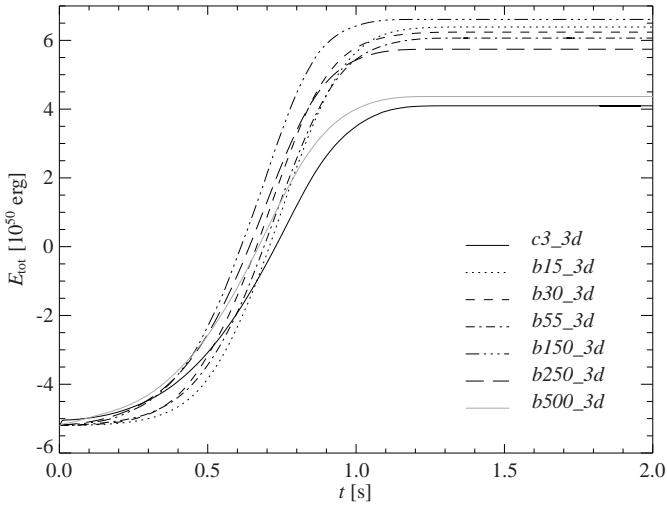}}
\caption{Total energies in the three-dimensional simulations.
  \label{3d_etot_fig}}
\end{figure}

With the three-dimensional implementation of our scheme, we performed
seven simulations with different initial flame configurations. The
three-dimensional analog of our standard test model \emph{c3\_2d},
termed \emph{c3\_3d}, served as comparison with two-dimensional runs
and with previous three-dimensional setups on other grid
geometries. Five ignition configurations with bubbles of a radius of
$5.5\,\mathrm{km}$ were chosen in a similar way as the setups for
the two-dimensional simulations of Sect.~\ref{twod_sect}. Again, a
Gaussian radial distribution was applied, with a dispersion of $\sigma
= 75 \, \mathrm{km}$. According to the number of igniting bubbles,
the models are denoted as \emph{b15\_3d, b30\_3d, b55\_3d, b150\_3d,}
and \emph{b250\_3d}. In yet another setup, \emph{b500\_3d}, we assumed
an equipartition over the radius instead of a Gaussian distribution. The
models are summarized in Table~\ref{3d_tab} and the distributions
of the ignition kernels in the multi-spot scenarios are depicted in
the left columns of
Figs.~\ref{inimod_1_3_fig} and \ref{inimod_4_6_fig}. Potentially, the
maximum distance of flame ignition is a relevant parameter since the
gravitational acceleration increases steeply in the inner part of the
WD. The values of the maximal ignition radii are given in
Table~\ref{3d_tab}. Fig.~\ref{3d_etot_fig} shows the energetic
evolution of the models.

\begin{table*}
\centering
\caption{Flame configurations and results of three-dimensional models.
\label{3d_tab}}
\setlength{\extrarowheight}{2pt}
\begin{tabular}{p{0.068\linewidth}rrrrrrr}
\hline\hline
model & 
\multicolumn{1}{p{0.095 \linewidth}}{number of ignition spots} & 
\multicolumn{1}{p{0.11 \linewidth}}{max. ignition\newline radius $[10^7 \, \mathrm{cm}]$} &
\multicolumn{1}{p{0.11 \linewidth}}{$M$(``Ni'') $[M_\odot]$} & 
\multicolumn{1}{p{0.11 \linewidth}}{$M$(``Mg'') $[M_\odot]$} & 
\multicolumn{1}{p{0.1 \linewidth}}{$M$(C,O) $[M_\odot]$} & 
\multicolumn{1}{p{0.1 \linewidth}}{$E_{nuc}$ $[10^{51} \, \mathrm{erg}]$} & 
\multicolumn{1}{p{0.1 \linewidth}}{$E_{tot}$ $[10^{50} \, \mathrm{erg}]$} \\
\hline
\emph{c3\_3d}   & (1) & 1.80 & 0.511 & 0.157 &0.738& 0.93 & 4.27 \\
\hline
\emph{b15\_3d}  & 15  & 1.68 & 0.657 & 0.158 &0.591& 1.16 & 6.62 \\
\emph{b30\_3d}  & 30  & 1.69 & 0.650 & 0.153 &0.603& 1.14 & 6.45 \\
\emph{b55\_3d}  & 55  & 1.47 & 0.633 & 0.165 &0.608& 1.13 & 6.29 \\
\emph{b150\_3d} & 150 & 1.80 & 0.667 & 0.167 &0.572& 1.18 & 6.81 \\
\emph{b250\_3d} & 250 & 1.83 & 0.613 & 0.164 &0.629& 1.09 & 5.94 \\
\hline
\emph{b500\_3d} & 500 & 1.80 & 0.526 & 0.162 &0.718& 0.96 & 4.57\\
\hline
\end{tabular}
\end{table*}

\subsection{Flame front evolution}

\begin{figure*}[t]
\centerline{
\includegraphics[width = \linewidth]
  {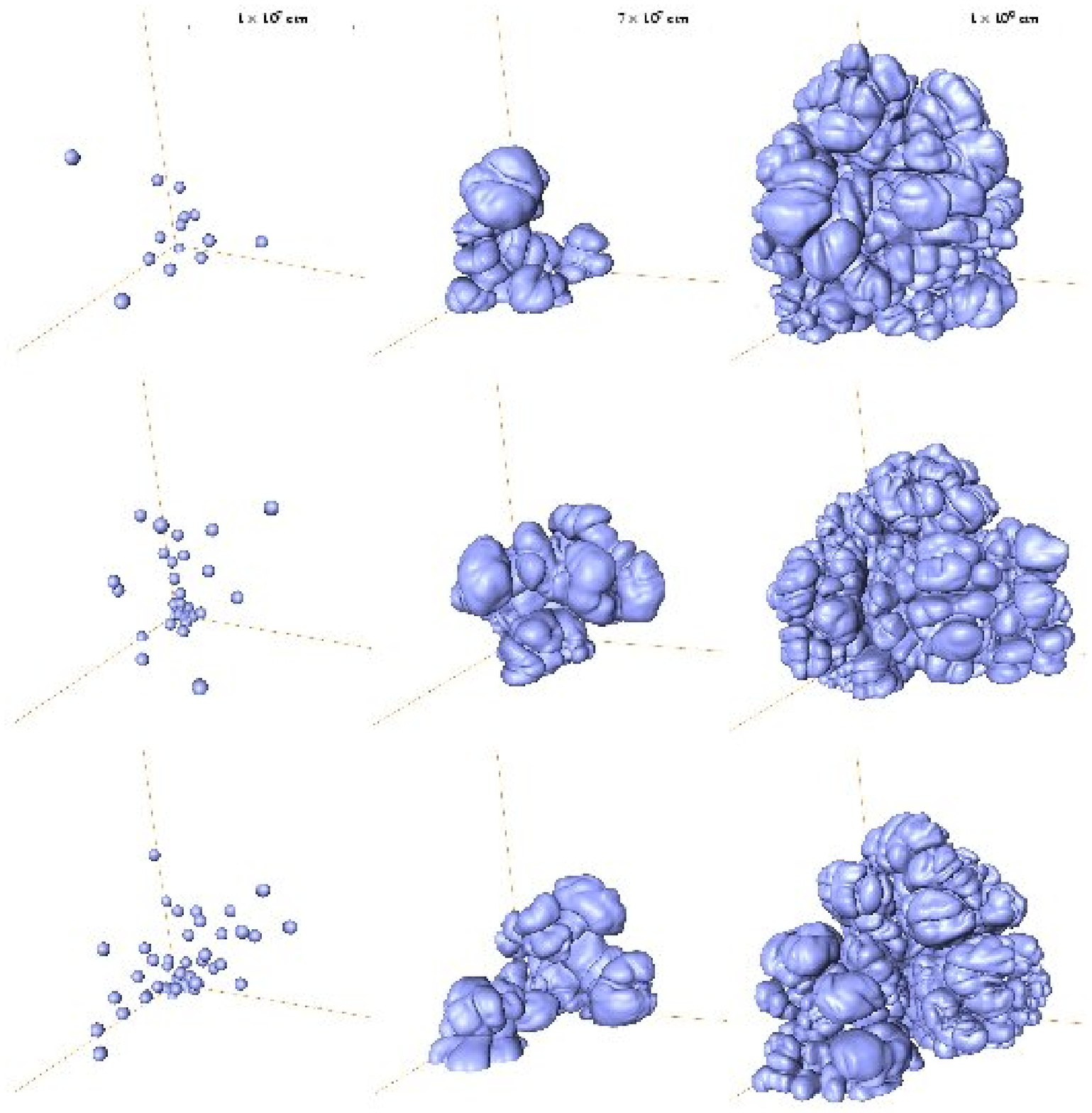}}
\caption{Evolution of the flame fronts in models \emph{b15\_3d} (top),
  \emph{b30\_3d} (middle), and  \emph{b55\_3d} (bottom), at times $t =
  0 \, \mathrm{s}$ (left column), $t =
  0.6 \, \mathrm{s}$ (middle column), and $t =
  2.0 \, \mathrm{s}$ (right column). The indicated length scale
  applies to the respective column.
  \label{inimod_1_3_fig}}
\end{figure*}

\begin{figure*}[t]
\centerline{
\includegraphics[width = \linewidth]
  {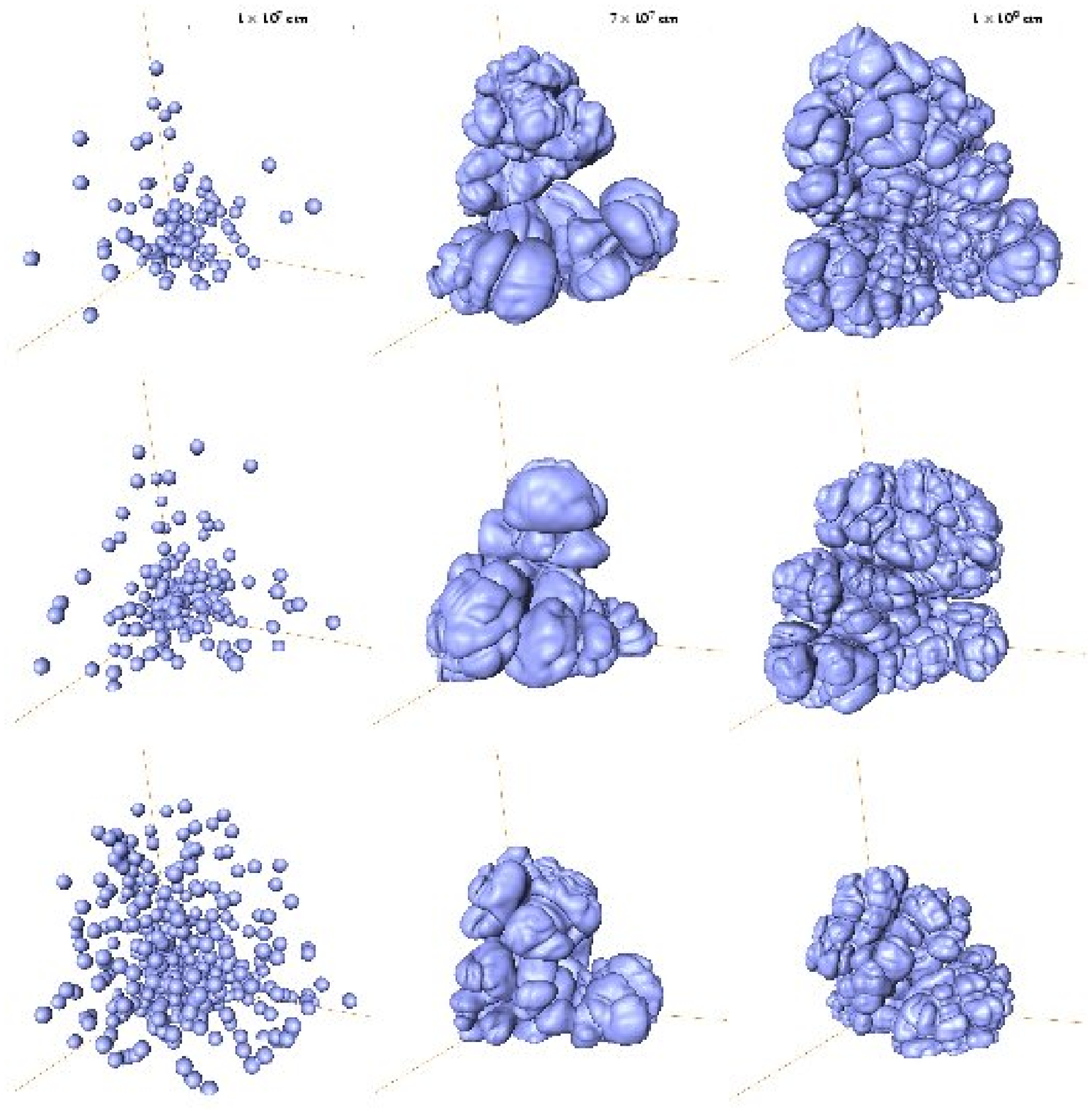}}
\caption{Evolution of the flame fronts in models \emph{b150\_3d} (top),
  \emph{b250\_3d} (middle), and  \emph{b500\_3d} (bottom), at times $t =
  0 \, \mathrm{s}$ (left column), $t =
  0.6 \, \mathrm{s}$ (middle column), and $t =
  2.0 \, \mathrm{s}$ (right column). The indicated length scale
  applies to the respective column.
  \label{inimod_4_6_fig}}
\end{figure*}

The evolutions of the flame fronts in our three-dimensional models with
multi-spot ignitions are shown in Figs.~\ref{inimod_1_3_fig} and
\ref{inimod_4_6_fig}. As expected, the basic features are the same in
all models and consistent with the findings of
\citet{reinecke2002d}. After ignition in multiple spots, the volumes of
the bubbles slowly increase due to the burning of the flame. In this
first stage, buoyancy-induced flotation of the bubbles is slow since
their dimension is small. Therefore the shear flows are weak and the
effects of turbulent flame wrinkling increase only gradually. This
corresponds to a slow increase of the released energies as can be seen
from Fig.~\ref{3d_etot_fig}. 

At $t \sim 0.4 \, \mathrm{s}$ the energy
generation rates increase drastically and peak around $0.6 \,
\mathrm{s}$. The flame fronts at this time are shown in the middle
column of Figs.~\ref{inimod_1_3_fig} and
\ref{inimod_4_6_fig}. 
We note here that the initially separated ignition spots have merged
into a single connected structure in agreement with
\citet{reinecke2002d} and \citet{roepke2005b}. As pointed out by
\citet{roepke2005b} (see also Fig.~5 of that publication), this is a
natural consequence of the flow field
that establishes around buoyantly rising burning bubbles. The stream
lines are directed around the forming mushroom-like structure and
converge behind it, where an upwards pointing flow emerges.
This flow drags underlying flame patches towards the bubble so
that the originally disconnected flame structures eventually merge.  
This effect has been ignored in the dimensional estimate of
Sect.~\ref{multi_sect}.

After $t \sim 1.5 \, \mathrm{s}$ the burning has
ceased in our models and the energies released in the explosion process
have reached their final values (cf.\ Fig.~\ref{3d_etot_fig}). The flame
fronts at $t = 2.0 \, \mathrm{s}$ are shown in the right columns of
Figs.~\ref{inimod_1_3_fig} and \ref{inimod_4_6_fig}. All models were
followed up to $t = 10.0 \, \mathrm{s}$ where homologous expansion is
reached with reasonable accuracy \citep{roepke2005c}. In the last
seconds the distributions of the ashes change only slowly in the
relaxation process.

From Figs.~\ref{inimod_1_3_fig} and \ref{inimod_4_6_fig} it is
obvious that the different ignition configurations lead to significant
variations in the flame evolutions. They are determined by the effects
mentioned in Sect.~\ref{multi_sect}.

The first effect is most relevant in cases of small numbers of
ignition spots. A sparse ignition may easily lead to anisotropies in
the evolving flame structure. Due to the randomness of flame kernel
locations the emerging large-scale flame structure is already
imprinted in the ignition seed. This is most obvious in model
\emph{b55\_3d}, where a preferential direction towards
observer (cf.\ Fig.\ref{inimod_1_3_fig}, lower row) is already present
in the ignition configuration and retained in the later evolution --
although somewhat moderated in the latest stages. Since there are only
very few ignition spots at large radii, i.e.\ subject to the large
gravitational accelerations, these tend to produce the largest
structures.
This effect is, however, less pronounced than in the
two-dimensional simulations. Thus, three-dimensional multi-spot models
give rise to a more robust flame evolution.

\begin{figure}[t]
\centerline{
\includegraphics[width = \linewidth]
  {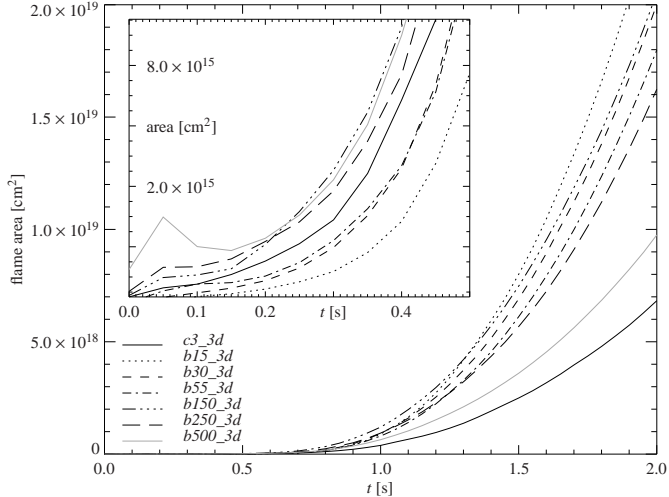}}
\caption{Flame surface area in different models as a function of
  time. The inset enlarges the evolution in the early stages.
  \label{fl_area_fig}}
\end{figure}

As expected, with increasing number of ignition spots the flames will
merge quickly due to burning and consequently flame surface is lost. To illustrate
this effect, the temporal evolution of an approximate measure of the
flame surfaces in the different models is plotted in
Fig.~\ref{fl_area_fig}. Obviously, they
vary considerably. As a consequence of the dense distribution of
initial flame bubbles in model \emph{b500\_3d}, the effect is most
pronounced there. From ignition to $t = 0.05 \, \mathrm{s}$ the flame
surface area increases rapidly, but it again decreases to a minimum
between $t = 0.1 \, \mathrm{s}$ and $0.2 \, \mathrm{s}$. The flame
surface destruction effect is also present in a less pronounced way in
the models \emph{b250\_3d} and \emph{b150\_3d} and even
\emph{b55\_3d}. In these, the flame
surfaces increase monotonically, but the slopes decrease between $t =
0.05 \, \mathrm{s}$ and $t = 0.15 \, \mathrm{s}$. In contrast, the
surface destruction is not noticeable in simulations starting from
sparse distributions of ignition spots. The slopes of flame surface
area evolution increase monotonically in models \emph{b15\_3d} and
\emph{b30\_3d}. Interestingly, the \emph{c3\_3d}
model corresponds to an intermediate case.

The comparison of the later flame area evolution is more
complicated. After $t \sim 1.0 \, \mathrm{s}$ the flame surface areas
of the models with sparse ignition seeds catch up with the
models ignited in denser initial flame configurations. One possible
explanation may be that in the latter, as soon as the initial flame
seeds have merged due to burning, a new structure emerges which is similar
to a single perturbed sphere. Since this large lump of
ashes is generated in a random way, seeds for larger and perhaps
dominating flame features may not emerge in a balanced distribution.
Thus, the flame evolution may again be hampered by large scale modes
as in the case of starting with sparse flame seeds and thus the loss in
flame surface area may never be recovered.

It should be noted that the radius inside of which the flame kernels are
ignited plays a minor role. In model \emph{b55\_3d} it was reduced by
about 10\% with little impact on the results.

\subsection{Global quantities}
\label{global_sect}

The evolutions of the total energies
in the models are plotted in Fig.~\ref{3d_etot_fig}. Values of the
total energies, the nuclear energy releases, and the masses of produced
iron group and intermediate mass nuclei and unburnt material at the
end of our simulations 
($t = 10.0 \, \mathrm{s}$) are included in
Table~\ref{3d_tab}. The masses of unburnt material in all models
are too large to be consistent with observations. This is partially
due to the fact that we stop burning artificially once the fuel
density drops below 
$10^7\,\mathrm{g}\,\mathrm{cm}^{-3}$. Realistically, burning should
continue to lower fuel densities converting more carbon/oxygen
material into intermediate mass elements thereby increasing the energy
release of the explosion \citep{roepke2005a}.

The most vigorous explosion resulted from model
\emph{b150\_3d} which released $1.18 \times 10^{51} \, \mathrm{erg}$
of nuclear energy producing $0.667 \, M_\odot$ of iron group elements and
$0.167 \, M_\odot$ of intermediate mass elements. The weakest explosion
was obtained with the \emph{c3\_3d} initial flame configuration, but
the model with most ignition spots, \emph{b500\_3d}, exploded almost
as weakly. These global characteristics are a natural consequence of
the flame area evolutions in the different models described
above. The fact that model \emph{b15\_3d} finally develops the largest
flame area is no contradiction. It catches up with the flame area of
model \emph{b150\_3d} at $t \sim 1.3 \, \mathrm{s}$, which is late in
the burning phase of the simulation. At this time the burning is
incomplete due to the low fuel densities in the expanded WD. It
terminates in intermediate mass nuclei and the energy
release is thus lower than in complete burning to nuclear statistical
equilibrium (NSE) in earlier stages.

The global characteristics of the explosion models confirm the
conjecture of Sect.~\ref{multi_sect}. The
explosion strength one can achieve with multiple ignition spots is
indeed limited and there exists an optimal flame configuration
with a finite number of initial flame kernels. Nevertheless, most
multi-spot ignition scenarios show a significantly increased energy
production as compared to the centrally ignited \emph{c3\_3d}
simulation. It is remarkable that -- in contrast to the
two-dimensional models -- most of the simulations reach total energies
within a narrow range of about $0.7\times 10^{50} \, \mathrm{erg}$
around $6.3 \times 10^{50} \, \mathrm{erg}$ (cf.\
Fig.~\ref{3d_etot_fig} ). This may be interpreted 
as a sign for convergence and robustness of the multi-spot ignition
model in its three-dimensional implementation. As far as the
global quantities are concerned, even the most energetic explosion
model \emph{b150\_3d} is not a clearly distinguished optimum, but
all models starting with 15 to 250 ignition spots per octant reach similar
explosion energies with some scatter due to the random choice of
initial flame locations.

\subsection{Distribution of elements in the ejecta}

\begin{figure*}[t]
\centerline{
\includegraphics[width = \linewidth]
  {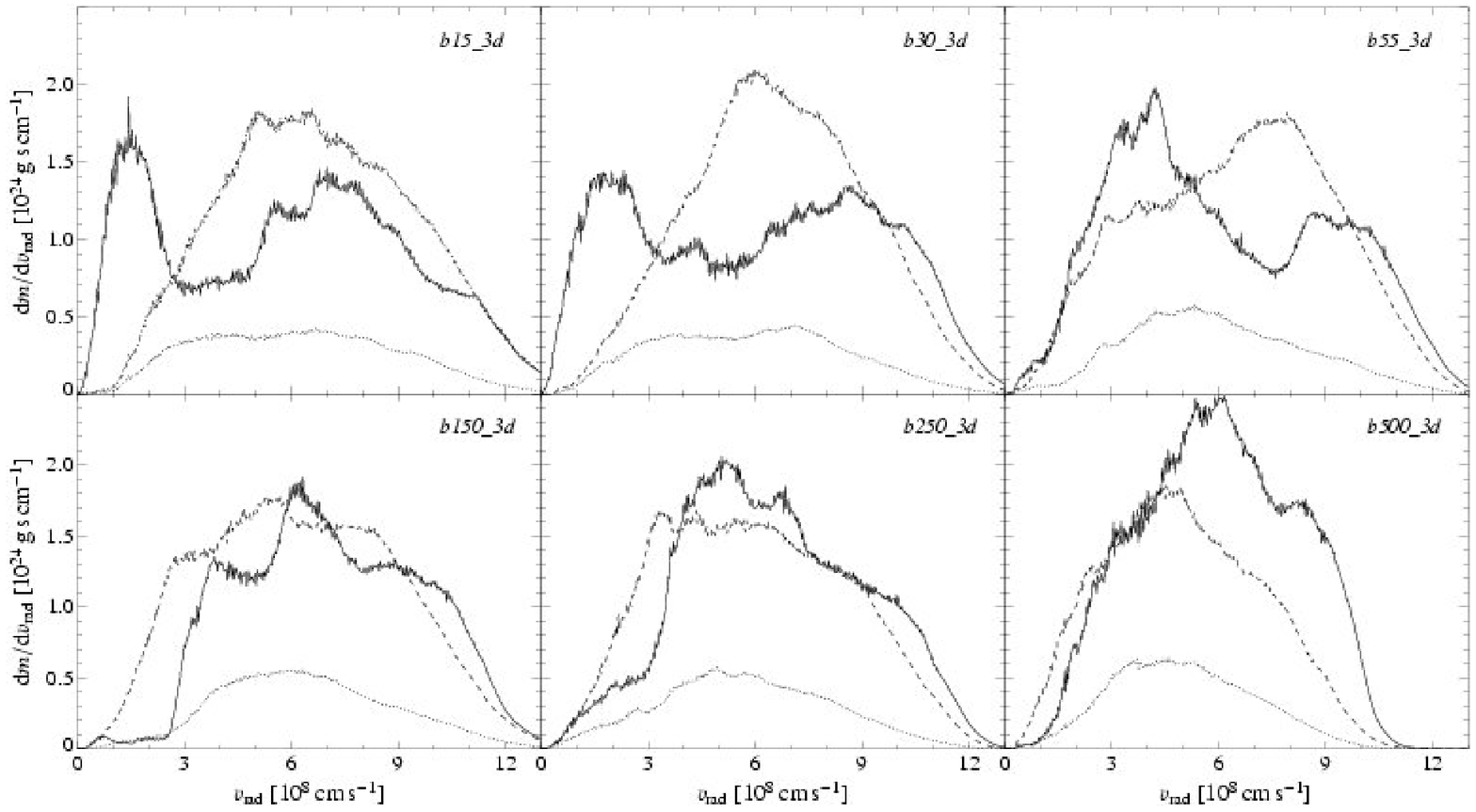}}
\caption{Distribution of the followed species in the three-dimensional
  models: carbon/oxygen fuel (solid) iron group elements (dashed),
  intermediate mass elements (dotted).
  \label{velprof_fig}}
\end{figure*}

Besides the global quantities as the production of energy, other
quantities can be compared with observational findings to judge the
validity of SN Ia explosion models. These are mainly the abundances
and distributions of various species in the explosion ejecta. 
Of course, most important is the radioactive isotope $^{56}$Ni, since
its decay powers the optical event, but also other isotopes 
contribute significantly to the shapes of spectra and light curves.

Constraints on abundances and distributions of elements can be obtained
from comparing synthetic light curves and spectra derived from the
models with observations. Alternatively, one can determine the
composition stratification of the ejecta by fitting model spectra to a
series of observed ones \citep{stehle2005a} and compare it with the
results of explosion models. Such detailed approaches are, however,
beyond the scope of the present paper.

Without nuclear postprocessing \citep[cf.][]{travaglio2004a} we can
only evaluate the cumulative abundances and distributions of iron group
elements, intermediate mass elements, and unburnt material (carbon and
oxygen). The produced masses of the former are given in
Table~\ref{3d_tab}. From comparison with the findings of
\citet{travaglio2004a} it seems likely that the amounts of $^{56}$Ni
synthesized in the most strongly exploding models will be close to $0.4 \,
M_\odot$. This value falls within the range of the expectations for
normal SNe Ia \citep{contardo2000a}, but rather on the low side.

The most interesting feature is the distribution of unburnt material
in velocity space plotted in Fig.~\ref{velprof_fig}. \citet{kozma2005a}
showed that spectra of the nebular phase put sharp constraints on
the models in this respect. A synthetic nebular spectrum derived from
a weakly exploding \emph{c3\_3d} model (but with a coarser flame
resolution than the model presented here due to differences in the
computational grids) featured strong emission lines originating from
oxygen at low velocities. This
effect might be an artifact of the peculiar (and unrealistic)
\emph{c3\_3d}-configuration, where the large buoyant structures are
already imprinted on the initial flame leading to strong downdrafts of
unburnt material. It may be suspected that
multi-spot ignition scenarios leading to different flame evolutions
alleviate the downdraft problem. This could be possible due to the more
complex initial flame structure which may be regarded as perturbed by
a wider range of modes. Moreover, due to the large flame surface area
that is achievable here, burning becomes more efficient and
increased amounts of the sinking fuel in the
downdrafts may be converted. 

Indeed, this conjecture is supported by our results (cf.\
Fig.~\ref{velprof_fig}). In particular, in model \emph{b250\_3d} the
unburnt material at low velocities is significantly reduced and the
composition is dominated here by iron group elements. With increasing
numbers of ignition spots a clear trend is noticeable from
Fig.~\ref{velprof_fig}. Sparse distributions of initial flames
still show an unburnt-material dominated composition at low velocities.
With increasing numbers of ignition spots this gradually changes into
an iron-group dominance. This effect is, however, not simply caused by the
fact that with denser distributions of flame kernels more material is
converted already in the ignition setup. The trend weakens with larger
numbers of initial flames (cf.\ models \emph{b250\_c3} and
\emph{b500\_c3} in Fig.~\ref{velprof_fig}). The reason is most likely
again the effect of flame kernels melting together shortly after
ignition, forming a large lump as described above. The emerging
random perturbations of this structure are likely to prevent a
balanced flame evolution similar to what is found in case of sparse
ignition spot distributions. Such flame propagation scenarios
naturally leave behind more unburnt material in central regions.

In contrast to the global quantities discussed in
Sect.~\ref{global_sect}, the distribution of the species in the
explosion ejecta clearly depends on the number of flame ignition
kernels. Models with $\sim$150 ignition spots per octant produce distributions
that are closest to observational expectations.

\section{Conclusions}
\label{concl_sect}

In the present study we have explored the possibilities of multi-spot
ignition scenarios of Type Ia supernova explosion models in a
systematic way. Owing to improvements of the computational code, it
was possible to achieve a high initial resolution of $1.74 \,
\mathrm{km}$ on computational grids with 256 cells per
dimension. By setting the initial bubble radii to $5.5 \, \mathrm{km}$, it
was possible to accommodate up to 500 flame kernels within a
radius of $\sim$$180 \, \mathrm{km}$ around the center
of the WD star. Due
to the random placement and overlap of the
bubbles, larger structures emerge providing a variety of sizes of
ignition seeds.  

Comparing the results of two-dimensional simulations with those of
three-dimensional models, we find significant differences in the flame
front evolutions starting from multiple ignition spots. In the former
setups, single outliers tend to dominate the evolution to a much
higher degree than in three dimensions.
We conclude from these findings that two-dimensional simulations are
inappropriate to explore initial flame configurations. Two-dimensional
studies can therefore only be a tool to study other effects and
compare simulations with the same fixed and carefully chosen initial
flame configuration, a possibility being the
\emph{c3\_2d}-model. Thus, although the general picture of our
two-dimensional simulations is similar to the findings of
\citet{livne2005a}, we disagree with their conclusions.

A simple estimate showed that the gain in explosion strength arising
from multi-spot ignition scenarios should be limited. This is
confirmed by a systematic study based on three-dimensional simulations
covering one octant of the star. Different ignition configurations with
15 up to 500 flame seeds were tested. We found the most vigorously
exploding model for 150 flame kernels per octant with a radial Gaussian
distribution inside a sphere of $180 \, \mathrm{km}$ around the center
of the WD. This agrees reasonably with our dimensional estimate in
Sect.~\ref{multi_sect}, given the simplifications made there. Setups
with increased resolutions
-- although allowing for smaller flame seeds -- would not profit from
much higher numbers of ignition spots. 

In terms of production of energy and iron group elements this
simulation came close to the \emph{b30}-model\footnote{Not to be
  confused with the \emph{b30\_2d} and \emph{b30\_2d} of the present study.} of
\cite{travaglio2004a}, which is a 
success of the hybrid grid implementation since it allows
results on an only $[256]^3$ cells computational grid  that previously
required $[768]^3$ cells. This novel approach will provide a powerful tool
to study the details of SN Ia deflagration models at moderate
computational expenses. The similarity with the \emph{b30}-model
suggests that the deflagration scenario of SNe Ia in its current
implementation (i.e. applying a multi-spot ignition with a
  Gaussian radial distribution within $\sim$$180 \, \mathrm{km}$
  around the center of the WD and ignoring burning below fuel
  densities of $10^7\, \mathrm{g}\, \mathrm{cm}^{-3}$) will not allow asymptotic kinetic
energies much higher than $\sim$$7\times 10^{50} \, \mathrm{erg}$ and the total
production of iron group elements will not greatly exceed $0.7 \,
M_\odot$. Since these values already fall in the range of
observational expectations \citep[cf.\ eg.][]{contardo2000a},
multi-spot ignition scenarios offer a way of modeling
``normal'' SNe Ia events. These global quantities seem to be rather robust
against variations in the number and configuration of flame
kernels. 

Comparing our results with those of \citet{garcia2005a} we note
agreement in some general results. Multi-spot ignition scenarios may
give rise to an explosion of the WD star. The evolution of the
large-scale features is similar in both studies. Our novel approach,
however, facilitated an increased resolution of the initial flame
configuration by a factor of 10 compared with
\citet{garcia2005a}. It was therefore possible to accommodate much
smaller flame kernels in greater numbers within the ignition
region. This may be the main
reason for disagreements in the conclusions (our more elaborate flame
description certainly contributes as well, but this is difficult to
disentangle). In contrast to \citet{garcia2005a} our results do not
indicate a convergence with ever increasing numbers of
flame seeds, but rather a maximum at a limited number. Moreover, our
initial flame bubbles seem to merge more
rapidly forming a connected flame structure. This gives rise to
differences in the distribution of species in the ejecta. The
clumpiness of ashes reported by \citet{garcia2005a} is less pronounced
in our simulations. 

In cases with larger numbers of ignition
spots, we find a much better exhaustion of the fuel in central
regions. In contrast to the global quantities, this effect is
sensitive to the number of ignition spots, indicating an
incompleteness of the current deflagration model of SNe
Ia. Potentially, an improved modeling of late phases of the burning
process \citep{roepke2005a} may be necessary, which possibly also
increases the explosion energy.

We conclude from our simulations that multi-spot ignition scenarios
may allow SN Ia models to achieve better agreement with observations,
although the effect is limited with regard to the number of igniting
flames. It thus does not lead to an arbitrary means of tuning the
models to higher energy release or nuclear conversion in the explosive
burning. Multi-spot ignition scenarios, however, may be capable of
reproducing ``normal'' SN Ia explosions.

\begin{acknowledgements}
We thank M.~Reinecke for helpful discussions. This work was supported
in part by the European Research Training Network ``The Physics of
Type Ia Supernova Explosions'' under contract HPRN-CT-2002-00303. The
research of JCN was supported by the Alfried Krupp Prize for Young
University Teachers of the Alfried Krupp von Bohlen und Halbach
Foundation. SEW acknowledges support from the NASA Theory Program
(NAG5-12036) and the DOE Program for Scientific Discovery
through Advanced Computing (SciDAC; DE-FC02-01ER41176).
\end{acknowledgements}

\end{document}